\documentclass[twocolumn,showpacs,preprintnumbers,amsmath,amssymb,superscriptaddress,pra,floatfix]{revtex4}
\usepackage{dcolumn}
\usepackage{bm,bbm}
\usepackage{graphicx,times}
\newcommand{\omegaz}{\omega_{0}}

\newcommand{\bmsigma}{\boldsymbol \sigma}

\newcommand{\bmA}{\boldsymbol A} 
\newcommand{\bmC}{\boldsymbol C}

\begin{document}
\title{Continuous variable entanglement dynamics in structured reservoirs}
\author{Ruggero Vasile}
\email{ruggero.vasile@utu.fi} \affiliation{Turku Center for Quantum
Physics, Department of Physics and Astronomy, University of Turku,
20014 Turun Yliopisto, Finland}
\author{Stefano Olivares}
\email{stefano.olivares@mi.infn.it}
\affiliation{Dipartimento di Fisica, Universit\`a degli Studi di Milano, I-20133 Milano, Italy}
\affiliation{CNISM UdR Milano Universit\`a, I-20133 Milano, Italy}
\author{Matteo G.~A.~Paris}
\email{matteo.paris@fisica.unimi.it}
\affiliation{Dipartimento di Fisica, Universit\`a degli Studi di Milano, I-20133 Milano, Italy}
\affiliation{CNISM UdR Milano Universit\`a, I-20133 Milano, Italy}
\affiliation{ISI Foundation, I-10133 Torino, Italy}
\author{Sabrina Maniscalco}
\email{sabrina.maniscalco@utu.fi} \affiliation{Turku Center for
Quantum Physics, Department of Physics and Astronomy, University of
Turku, 20014 Turun Yliopisto, Finland}
\date{\today}
\begin{abstract}
We address the evolution of entanglement in bimodal continuous
variable quantum systems interacting with two independent structured
reservoirs. We derive an analytic expression for the entanglement of
formation without performing the Markov and the secular
approximations and study in details the entanglement dynamics for
various types of structured reservoirs and for different reservoir
temperatures, assuming the two modes initially excited in a
twin-beam state.  Our analytic solution allows us to identify three
dynamical regimes characterized by different behaviors of the
entanglement: the entanglement sudden death, the non-Markovian
revival and the non-secular revival regimes.  Remarkably, we find
that, contrarily to the Markovian case, the short-time
system-reservoir correlations in some cases destroy quickly the
initial entanglement even at zero temperature.
\end{abstract}
\pacs{03.67.Mn,03.65.Yz}
\maketitle
\section{Introduction}
Entanglement is an essential resource for quantum computation and
communication protocols \cite{NieChu}. However, this fundamental
quantum property is also fragile: the unavoidable interaction
of quantum systems with their external environment leads to the
irreversible loss of both quantum coherence (decoherence) and
quantum correlations in multi-partite systems
\cite{Breuer,weiss,ZurekRev2003}.
\par
A crucial requirement for a physical system to be of interest for
quantum technologies is that the survival time of entanglement is
longer than the time needed for information processing. Therefore it
is important to develop a deep and precise understanding not only of
the mechanisms leading to decoherence and entanglement losses but
also of the dynamical features of these phenomena. Moreover, in view
of recent developments in reservoir engineering techniques
\cite{engineeredres,Zoller08}, it is interesting to investigate
situations in which decoherence and disentanglement can be
controlled, for example through a precise and accurate tuning of
system and environment parameters.
\par
In this paper we consider the entanglement dynamics in noisy
continuous variable (CV) quantum systems \cite{Braunstein}. More
specifically we focus our attention on a system of two
non-interacting quantum harmonic oscillators bilinearly coupled to
two independent structured reservoirs at temperature $T$. Our aim is
to study the time evolution of the entanglement between the two
oscillators for different temperature regimes, different
system-reservoir parameters and different reservoir spectra. Rather
than limiting ourself to present a plethora of dynamical behaviors,
we will try to identify general features in order to single out
universal properties of the disentanglement process, namely those
properties that do not depend either on the specific model of
reservoir chosen or on the specific initial value of the
entanglement. Moreover, we also compare the differences in the
dynamics arising from different spectral distributions of the
reservoir in order to identify those physical contexts leading to
stronger or weaker entanglement losses.
\par
During the last decade numerous works dealing with losses and
decoherence in bimodal CV quantum systems have appeared in the
literature. In order to describe analytically the dynamics of such
an open quantum system, approximations such as the Born-Markov and
the secular (or rotating wave) approximations are typically
performed
\cite{Xiang2008,Dodd2004,DoddHall2004,Hiro2001,PraBec2004,SerIllPar2004}.
The Markovian approximation basically consists in neglecting the
short-time correlations between system and reservoir arising because
of the structure of the reservoir spectrum. This approximation is
often performed together with the weak system-reservoir coupling
assumption, also known as Born approximation. The Born and the
Markov approximations are generally related. Indeed every time the
coupling between the system and the environment is strong, and
therefore the Born approximation is not appropriate, also the
Markovian approximation cannot be consistently used. However, there
exist situations of weak system-reservoir coupling and structured
environment, where the system-reservoir correlations persist long
enough to require a non-Markovian treatment, even in the weak
coupling limit. In this paper we focus on these cases.
\par
We also investigate the validity of the secular approximation (i.e.
neglecting the counter-rotating terms in the Hamiltonian) and we find
that, even for weak couplings, a correct description of the short-time
dynamics must take into account the nonsecular terms. In more detail,
the validity of the secular approximation sensibly depends both on the
reservoir temperature and on the system-reservoir parameters.
\par
Non-Markovian studies of bimodal CV quantum systems in a
common reservoir have shown the existence of three different
dynamical phases of the entanglement in the long time-limit, namely
the sudden death, sudden death and revivals, and no-sudden death
phases \cite{Paz08,Paz09}. These phases depend not only on the
system-reservoir parameters but also on the properties of the
spectrum. In this paper we consider the case of two independent
reservoirs and find a similar division in dynamical phases or
regimes, namely, the entanglement sudden death (ESD), the
non-Markovian revival (NMRev) and the non-secular revival (NSRev)
regimes. In our system, however, the
no-sudden death phase appears only at zero temperature and under
very specific conditions. Moreover, we have discovered that the
revivals may be due to two different physical mechanisms, the
non-Markovian finite reservoir memory or the presence of the non
secular terms.
\par
Recent literature on non-Markovian CV dynamics, in the common
reservoir scenario, includes Refs. \cite{Shiokawa,An}, while the
independent reservoirs case was considered using a phenomenological
approach in Ref. \cite{Masashi} and using a numerical approach in
Ref. \cite{LiuGoan,JunHong2009}. In this paper, we extend in several
directions the results we have obtained in Ref. \cite{MaOliPa},
where we limited our study to the high-$T$ Ohmic reservoir in the
secular approximation.  Here we solve the Master equation for our
system without performing the secular approximation and investigate
quantitatively the entanglement dynamics using an analytic
expression for the evolution of the entanglement of formation (EoF)
\cite{EoF,G:03,EoF2008}. We assume the two oscillators initially
excited in a twin-beam state (TWB, sometimes also referred to as
two-mode squeezed states) and consider Ohmic, sub-Ohmic and
super-Ohmic reservoirs at any temperature.
\par
The paper is organized as follows. In Sec. \ref{s:ME} we introduce the
physical system, the Master equation and its general solution
through the characteristic function approach. In Sec. \ref{s:EDynGS} we review
some preliminary concepts about two-mode Gaussian states and we
present the general solution of the Master equation with an initial
Gaussian state. We also introduce the TWB states, the concept of EoF
for two-mode CV Gaussian states, and the types of reservoir spectra
considered in the paper. In Sec. \ref{s:ValSA} we present a detailed
investigation on the validity of the secular approximation in our
model. In Sec. \ref{sec:ED} we discuss the dynamics of entanglement and
analyze the three emerging dynamical regimes: ESD, NMRev and NSRev.
Moreover, we give specific examples of the dynamics of the EoF
focusing on the high-$T$ and low-$T$ regimes. Finally, in Sec. \ref{s:concl}
we discuss and summarize our results, presenting conclusions and
future prospectives.
\section{The Master Equation}\label{s:ME}
We consider a system of two identical non-interacting quantum
harmonic oscillators, each of them coupled to its own bosonic
structured reservoir. The total Hamiltonian can be written as
\begin{eqnarray}
H&=&\sum_{j=1,2}\hbar\omega_0 a^{\dag}_j
a_j+\sum_{j=1,2}\sum_k\hbar\omega_{jk}
b^{\dag}_{jk} b_{jk} \nonumber \\
&+&\sum_{j=1,2}\sum_k
\gamma_{jk}(a_j+a_j^{\dag})(b_{jk}+b_{jk}^{\dag}),
\end{eqnarray}
with $\omega_0$ the oscillators frequency,  $\omega_{1k}$ and
$\omega_{2k}$ the frequencies of the reservoirs modes,  $a_{j}$
($a_{j}^{\dag}$) and $b_{jk}$ ($b_{jk}^{\dag}$) the annihilation
(creation) operators of the system and reservoirs harmonic
oscillators, respectively, and $\gamma_{jk}$ the coupling between
the $j$-th oscillator and the $k$-th mode of its environment. In the
following we assume that the reservoirs have the same spectrum and
are equally coupled to the oscillators.
\par
Since we are interested in the dynamics of the two oscillators only,
we adopt a density matrix approach through the following local in
time Master equation \cite{HuPazZhang}
\begin{equation}\begin{split}\label{HuPaZang}
&\dot{\rho}(t)=\sum_j\frac{1}{i\hbar}[H_j^{0},\rho(t)]-\Delta(t)
[X_j,[X_j,\rho(t)]]+\\
&+\Pi(t)[X_j,[P_j,\rho(t)]]+
\frac{i}{2}r(t)[X_j^2,\rho(t)]+\\
&-i\gamma(t)[X_j,\{P_j,\rho(t)\}],
\end{split}\end{equation}
where $\rho(t)$ is the reduced density matrix, $H_j^{0}$
is the free Hamiltonian of the $j$-th oscillator, and
$X_j=(a_j+a_j^{\dag})/\sqrt{2}$ and
$P_j= i (a_j^{\dag}-a_j)/\sqrt{2}$
are the quadrature operators. The effect of the
interaction with the reservoirs is contained in the time-dependent
coefficients of Eq. \eqref{HuPaZang}. The quantities $\Delta(t)$ and
$\Pi(t)$ describe diffusion processes, $\gamma(t)$ is a damping term and
$r(t)$ renormalizes the free oscillator
frequency $\omega_0$.
\par
It is worth noting that the Master equation \eqref{HuPaZang} is
exact, since neither the Born-Markov approximation nor the secular
approximation have been performed. The time dependent coefficients
can be expressed as power series in the system-reservoir coupling
constant. For weak couplings one can stop the expansion to second
order and obtain analytic solutions  for the coefficients. In the
case of reservoirs in thermal equilibrium at temperature $T$,
characterized by a spectral density $J(\omega)$, these expressions
read
\begin{subequations}\label{Coeff1}
\begin{align}
\Delta(t)&=\alpha^2\!\!\int_0^t\!\! ds \! \int_0^{+\infty}\!\!\!\!\!\!\!\!
\!d\omega
J(\omega)[2N(\omega)+1]\cos(\omega s)\cos(\omega_0 s), \\
\Pi(t)&=\alpha^2\!\!\int_0^t\!\! ds \! \int_0^{+\infty}\!\!\!\!\!\!\!\!
\!d\omega
J(\omega)[2N(\omega)+1]\cos(\omega s)\sin(\omega_0 s), \\
\gamma(t)&=\alpha^2\!\!\int_0^t\!\! ds \! \int_0^{+\infty}\!\!\!\!\!\!\!\!
\!d\omega
J(\omega)\sin(\omega s)\sin(\omega_0 s), \\
r(t)&=\alpha^2\!\!\int_0^t\!\! ds \!
\int_0^{+\infty}\!\!\!\!\!\!\!\! \!d\omega J(\omega)\sin(\omega
s)\cos(\omega_0 s),
\end{align}
\end{subequations}
where $N(\omega)=[\exp(\hbar\omega/k_BT)-1]^{-1}$ is the mean
number of photons with frequency $\omega$, and $\alpha$ is the dimensionless
system-reservoir coupling constant.
\par
By using the characteristic function approach \cite{Intra2003}, the
solution of the Master equation \eqref{HuPaZang} may be written as
\begin{equation}\begin{split}\label{CharSol}
\chi_t(\Lambda)=\mbox{} &e^{-\Lambda^T
[\bar{W}(t)\oplus\bar{W}(t)]\Lambda}\\
&\times\chi_0(e^{-\Gamma(t)/2}[R^{-1}(t)\oplus R^{-1}(t)]\Lambda),
\end{split}\end{equation}
where $\chi_t(\Lambda)$
is the characteristic function at time $t$, $\chi_0$ is the
characteristic function at the initial time $t=0$,
$\Lambda = (x_1,p_1,x_2,p_2)$ is the two-dimensional phase space
variables vector, $\Gamma(t)=2\int_0^t\gamma(t')dt'$, and
$\bar{W}(t)$ and $R^{-1}(t)$ are $2 \times 2$ matrices.
The former matrix is given by
\begin{equation}
\bar{W}(t)=e^{-\Gamma(t)}[R^{-1}(t)]^T W(t)R^{-1}(t),
\end{equation}
while the latter one, $R(t)$, contains rapidly oscillating terms.
In the weak coupling limit $R(t)$ takes the form
\begin{equation}
R(t)=\left(
      \begin{array}{cc}
        \cos\omega_0 t & \sin\omega_0 t \\
        -\sin\omega_0 t & \cos\omega_0 t \\
      \end{array}
    \right).
\end{equation}
Finally, $W(t)=\int_{0}^{t}e^{\Gamma(s)}\bar{M}(s)ds$ with
$\bar{M}(s)=R^{T}(s)M(s)R(s)$ and
\begin{equation}
\label{eq:Mt} M(s)=\left(
       \begin{array}{cc}
         \Delta(s) & -\Pi(s)/2 \\
         -\Pi(s)/2 & 0 \\
       \end{array}
     \right).
\end{equation}
The coefficient $r(t)$ does not appear explicitly in the
characteristic function solution because its contribution is
negligible in the weak coupling regime \cite{Intra2003}.
The characteristic function approach of Ref. \cite{Intra2003} is
equivalent to other methods of solution of the Master equation
\eqref{HuPaZang}, as the Feynman-Vernon influence functional
technique \cite{Feynman1963}. In this paper we use the former one
because it allows to obtain an analytic solution in the weak
coupling limit.
\section{Entanglement dynamics for Gaussian states}\label{s:EDynGS}
In this section we derive the explicit analytic solution for
the characteristic function in the weak coupling limit already
obtained in \cite{Intra2003}. Remarkably, the evolution induced
by the Master Equation \eqref{HuPaZang} corresponds to a Gaussian
map, i.e. an initial Gaussian state maintains its character.
It is thus possible to obtain the expression of the covariance
matrix at time $t$ and then evaluate the EoF at any time for the
two modes initially excited in a TWB state.
We also introduce the classes of spectral densities considered in
the paper and show how the form of the time dependent coefficients
ruling the dynamics can be simplified in the non-Markovian time
scale.
\subsection{Analytic solution in the weak coupling limit}
Let us consider two-mode Gaussian states, i.e.,  those states
characterized by a Gaussian characteristic function
\begin{equation}
\chi_0(\Lambda)=\exp\biggl\{-\frac{1}{2}\Lambda^{T}
\sigma_0\Lambda-i\Lambda^{T}\bar{\mathbf{X}}_{in}\biggl\}.
\end{equation}
We indicate with $\sigma_0$ the initial  covariance matrix
\begin{equation}\label{CovMatzero}
\sigma_0=\left(
          \begin{array}{cc}
            \mathbf{A_0} & \mathbf{C_0} \\
            \mathbf{C^T_0} & \mathbf{B_0} \\
          \end{array}
        \right),
\end{equation}
where $\mathbf{A_0}= a\,{\mathbbm 1}$, $\mathbf{B_0}=b\,{\mathbbm 1}$,
$\mathbf{C_0}={\rm Diag}(c_1,c_2)$, with $a$,$b>0$ and $c_1$, $c_2$
real numbers, and $\mathbbm 1$ the $2\times 2$ identity matrix. Moreover,
\begin{equation}
\bar{\mathbf{X}}_{in}={\rm Tr}[\rho(0)(X_1,P_1,X_2,P_2)^T]\,.
\end{equation}
If $c_1=c_2=0$ the initial covariance matrix is block diagonal and
the corresponding state is separable. Since each oscillator
only interacts with its own environment, an initial separable state
remains separable during all the evolution. For initial entangled states,
however, the entanglement dynamics will in general depend on the initial
value of the entanglement and on reservoir properties such as the spectral
distribution, the temperature and the coupling constants.
\par
Since the evolution maintains the Gaussian character the evolved state
is a two-mode Gaussian state with mean and covariance matrix given by
\begin{align}
\bar{\mathbf{X}}_t&=e^{-\Gamma(t)/2}(R\oplus R)
\bar{\mathbf{X}}_{in} \\
\label{CovMatT}
\sigma_t&=e^{-\Gamma(t)}(R\oplus R)\sigma_0(R\oplus
R)^T+2(\bar{W}_t\oplus\bar{W}_t),
\end{align}
Using Eqs.~\eqref{CharSol}--\eqref{eq:Mt} we obtain
\begin{equation}\begin{split}\label{At}
\bar{W}_t=\mbox{}&e^{-\Gamma(t)}\int_0^t e^{\Gamma(s)}\biggl[
\frac{\Delta(s)}{2}\mathbbm{1}+\frac{\Delta(s)}{2}\mathbf{C}_2(t-s)\\
&-\frac{\Pi(s)}{2}\mathbf{S}_2(t-s)\biggl]ds,
\end{split}\end{equation}
where
\begin{equation}
\mathbf{C}_2(t)=\left(
      \begin{array}{cc}
        \cos 2\omega_0 t & -\sin 2\omega_0 t \\
        -\sin 2\omega_0 t & -\cos 2\omega_0 t \\
      \end{array}
    \right),
\end{equation}
\begin{equation}
    \mathbf{S}_2(t)=\left(
      \begin{array}{cc}
        \sin 2\omega_0 t & \cos 2\omega_0 t \\
        \cos 2\omega_0 t & -\sin 2\omega_0 t \\
      \end{array}
    \right).
\end{equation}
The covariance matrix at time $t$ is given by
\begin{equation}\label{CovMatT2}
\bmsigma_{t} = \left(
\begin{array}{c | c}
\bmA_{t} & \bmC_{t} \\ \hline \bmC_{t}^{T} & \bmA_{t}
\end{array}
\right),
\end{equation}
with
\begin{align}
\bmA_{t} ={}& \bmA_{0} e^{-\Gamma}\nonumber \\
& + \left(
\begin{array}{cc}
\Delta_{\Gamma} + (\Delta_{\rm co} - \Pi_{\rm si}) &
-(\Delta_{\rm si} - \Pi_{\rm co}) \\
-(\Delta_{\rm si} - \Pi_{\rm co}) & \Delta_{\Gamma} - (\Delta_{\rm
co} - \Pi_{\rm si})
\end{array}
\right),
\end{align}
and
\begin{align}
\bmC_{t} = \left(
\begin{array}{cc}
c\, e^{-\Gamma}\, \cos(2\omega_0 t) &
c\, e^{-\Gamma}\, \sin(2\omega_0 t) \\
c\, e^{-\Gamma}\, \sin(2\omega_0 t) & -c\, e^{-\Gamma}\,
\cos(2\omega_0 t)
\end{array}
\right),
\end{align}
where we have introduced the function
\begin{equation}\label{CoeffSep}
\Delta_{\Gamma}(t)=e^{-\Gamma(t)}\int_0^t e^{\Gamma(s)}\Delta(s)ds
\end{equation}
and the secular coefficients
\begin{subequations}
\label{SecCoeff}
\begin{align}
&\Delta_{co}(t)=e^{-\Gamma(t)}\int_0^t
e^{\Gamma(s)}\Delta(s)\cos[2\omega_0(t-s)]ds,\\
&\Delta_{si}(t)=e^{-\Gamma(t)}\int_0^t e^{\Gamma(s)}\Delta(s)\sin[2\omega_0(t-s)]ds,\\
&\Pi_{co}(t)=e^{-\Gamma(t)}\int_0^t
e^{\Gamma(s)}\Pi(s)\cos[2\omega_0(t-s)]ds,\\
&\Pi_{si}(t)=e^{-\Gamma(t)}\int_0^t
e^{\Gamma(s)}\Pi(s)\sin[2\omega_0 (t-s)]ds.
\end{align}\end{subequations}
The explicit analytic expression of the coefficients above depends on
both the reservoir spectral density and the temperature.
\par
To further simplify the solution it is common to perform the
so-called secular approximation. This approximation amounts at
neglecting rapidly oscillating terms in the solution of the Master
equation. In our case this means to assume that the coefficients
\eqref{SecCoeff} average out to zero. Stated another way, the
secular solution is a coarse-grained expression of the exact one. In
the next section we will critically examine the validity of the
secular approximation and derive the conditions of validity for
different reservoir spectra and system-reservoir parameters.
\par
From now on we focus on TWB states, i.e.,
a set of Gaussian states whose covariance matrix \eqref{CovMatzero}
has $a=b=\cosh(2r)/2$ and $c_1=-c_2=\sinh(2r)/2$,
with $r > 0$ the squeezing parameter. Being pure states, their
amount of entanglement is given by the entropy of entanglement
$E_0(r) = 2[\cosh^2r \, \ln (\cosh r) - \sinh^2r \, \ln (\sinh r)]$
and, hence, it increases for increasing values of $r$.
\subsection{Entanglement of Formation}

A convenient and useful way of looking at the entanglement evolution
in CV systems is by means of the EoF \cite{EoF,G:03}.
This quantity corresponds to the minimal amount of
entanglement of any ensemble of pure bipartite states realizing the
given state. In general it is not a simple task to derive an expression of
the EoF for arbitrary states. Recently its expression for an
arbitrary bimodal Gaussian state has been obtained in Ref. \cite{EoF2008}.
\par
We assume here that the initial state is a symmetric bipartite
Gaussian state with covariance matrix given by Eq. \eqref{CovMatzero}.
As we mentioned above, when
this state interacts with two identical independent reservoirs,
the Gaussian character is preserved and the evolved covariance
matrix is given by Eq. \eqref{CovMatT2}. Due to the symmetry of the
evolved state, the EoF is given by \cite{G:03}
\begin{equation}\label{EoF}
E_{F} = (x_m + \mbox{$\frac12$}) \ln(x_m +\mbox{$\frac12$}) - (x_m -
\mbox{$\frac12$}) \ln(x_m - \mbox{$\frac12$}),
\end{equation}
with $x_m = (\tilde\kappa_{-}^{2} + 1/4)/(2 \tilde\kappa_{-})$,
$\tilde\kappa_{-}=\sqrt{(a_n - c_{+})(a_n - c_{-})}$ being the minimum
symplectic eigenvalue of the CM $\bmsigma_{t}$, and
\begin{align}
a_n &= \sqrt{I_1},\\
c_{\pm} &= \sqrt{\frac{I_1^2+I_3^2-I_4 \pm \sqrt{(I_1^2+I_3^2-I_4)^2
- (2 I_1I_3)^2}} {2 I_1}}, \label{c:pm}
\end{align}
where $I_1 = \det[\bmA_t]$, $I_3 = \det[\bmC_t]$ and $I_4 =
\det[\bmsigma_t]$ are the symplectic invariants of $\bmsigma_t$.
Inserting Eqs. \eqref{CovMatT2}-\eqref{SecCoeff}  into Eqs.
\eqref{EoF}--\eqref{c:pm} one obtains the analytic expression of the
EoF for our system.
\subsection{Modeling the reservoir}
In order to obtain explicit expressions for the EoF we need to specify the
properties of the bosonic reservoirs. We consider environments in
thermal equilibrium at temperature $T$ and we focus on the
following class of Ohmic-like spectral
distributions with an exponential cut-off function
\begin{equation}\label{Reservoirs}
J_s(\omega)=\omega_c\biggl(\frac{\omega}{\omega_c}\biggl)^s
e^{-\omega/\omega_c},
\end{equation}
where $\omega_c$ is the cut-off frequency. The case $s=1$
corresponds to an Ohmic reservoir spectrum, characterized by a
linear dependence on the frequency for $\omega\ll \omega_c$. For $s>1$
the spectrum is known as super-Ohmic while $s<1$ describes a
sub-Ohmic spectral distribution. For the sake of concreteness, in
the following we consider the $s=3$ super-Ohmic and the $s=1/2$
sub-Ohmic cases. A more detailed discussion about the properties of
these spectral distributions can be found, e.g., in Ref. \cite{Paa}.
\par
A closed form for the expressions of the time-dependent coefficients
given in Eq. \eqref{Coeff1} can be obtained in the high-$T$ and
zero-$T$ limits, i.e., for
$2N(\omega)+1\approx\frac{2k_BT}{\hbar\omega}$ and
$2N(\omega)+1\approx 1$, respectively (See Appendix A). Therefore we
focus on these two regimes. Inserting now the spectral distributions
of Eq. \eqref{Reservoirs}, with $s=1$, $s=1/2$, and $s=3$, into Eqs.
\eqref{Coeff1} allows to determine the analytic form of the
time-dependent coefficients. We notice that, in all three cases,
after a time $t \approx \tau_c = \omega_c^{-1}$, the coefficients
attain their Markovian stationary values and the system behaves
according to the predictions of the Markovian theory.  Here we are
particularly interested in the non-Markovian short-time dynamics,
and therefore we will focus on times $t \le \tau_c$. In this time
interval, and in the weak coupling limit, we can expand the
exponential terms appearing in Eqs. \eqref{CoeffSep} and
\eqref{SecCoeff}  in Taylor series. For example Eq. \eqref{CoeffSep}
becomes
\begin{equation}\begin{split}\label{TaylorDelta}
&\Delta_{\Gamma}(t)\simeq\int_0^t\Delta(s)ds-\Gamma(t)\int_0^t\Delta(s)ds+\\
&+\int_0^t \Gamma(s)\Delta(s)ds+O(\alpha^4).
\end{split}\end{equation}
Since $\Delta(t) \propto \alpha^2$ and  $\Gamma(t) \propto
\alpha^2$, in the weak coupling limit ($\alpha\ll 1$) and for short
non-Markovian times the first term dominates and hence it is the
only one that will be retained.
\section{On the validity of the secular approximation}\label{s:ValSA}
In this section we question the validity of the secular
approximation by comparing the entanglement dynamics with or without
the secular terms \eqref{SecCoeff}. As we will see, in general, the
secular terms do influence the behavior of the entanglement in the
short non-Markovian time scale. Depending on the value of certain
parameters, however, the secular approximation in some cases turns
out to give a good description of the dynamics. We have identified
as main parameters influencing the time evolution the reservoir
temperature, the parameter $x=\omega_c/\omegaz$, and the form of the
reservoir spectrum. We will discuss the effect of these parameters
separately in the following three subsections dealing with the
dynamics for high-$T$ reservoirs, $T=0$ reservoir, and with a
comparison between different reservoir spectra. We will also
consider if and how the validity of the secular approximation
depends on the initial state of the system and, in particular, on
the initial squeezing parameter $r$ of TWBs.
\par
We recall that in this paper we will focus only on the dynamics of
entanglement. Different observables of the system may show
different sensitivity to the secular approximation. Indeed we know
that there exist a class of observables, e.g., the energy of the
system oscillators, that are not influenced at all by this
approximation \cite{Intra2003}.
\subsection{High-temperatures regime}
We begin addressing the high-temperature limit
$k_BT\gg\hbar\omega_0,\hbar\omega_c$, i.e., when the classical
thermal energy $k_BT$ is much larger than the typical energies
exchanged in our system. In the following analysis we choose a
temperature such that $k_BT/\hbar\omega_c=100$, thus we can examine
scenarios in which $x = \omega_c / \omega_0\geq0.1$.
\par
We start analyzing the differences in the EoF evolution between the
secular result and the exact one in the case of an Ohmic
distribution and $x=10$.
\begin{figure}[h!]
\begin{center}
\includegraphics[width=0.49\columnwidth]{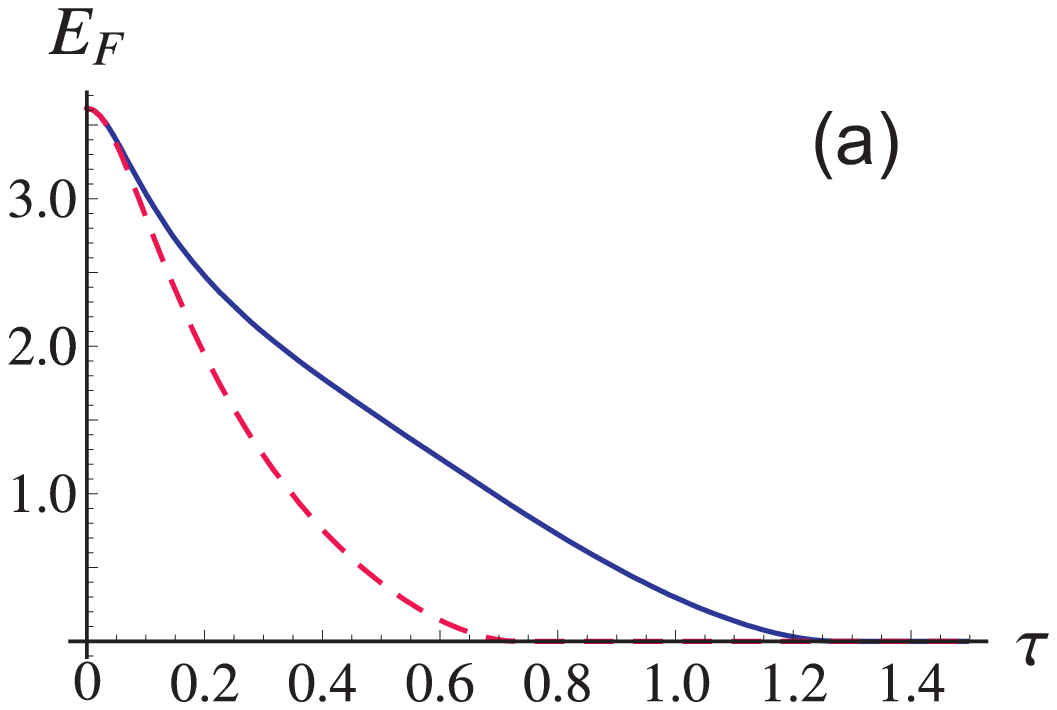}
\includegraphics[width=0.49\columnwidth]{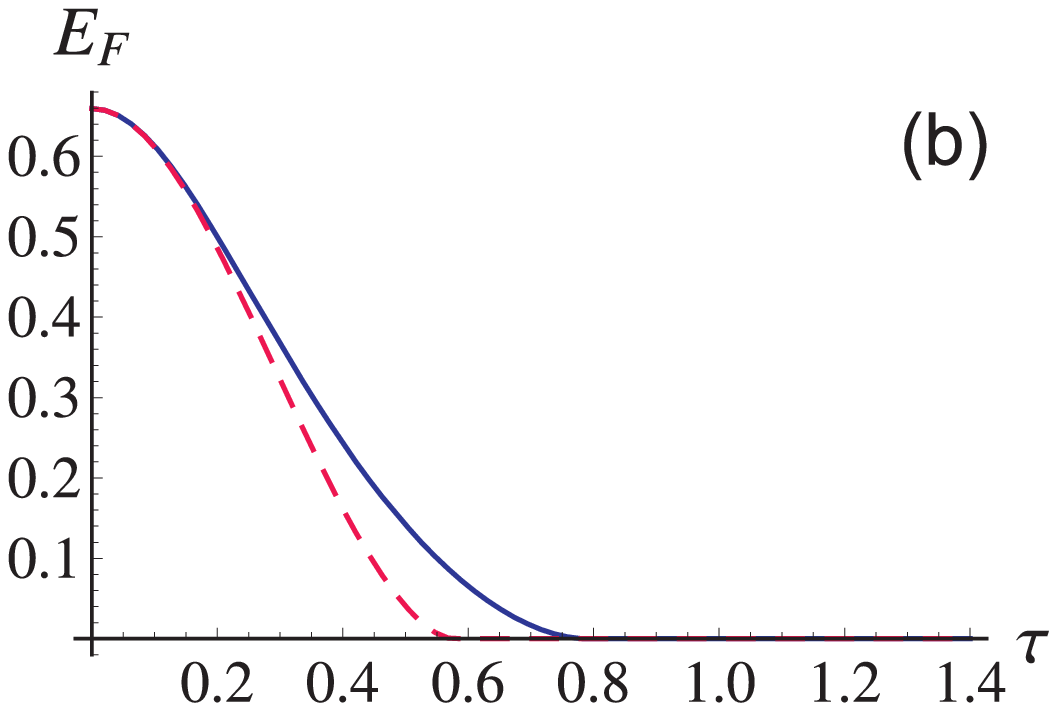}
\end{center}
\caption{(Colors online) Comparison between the exact EoF
dynamics (solid blue line) and the secular approximate
dynamics (dashed red line) as a function of $\tau=\omega_c t$,
with $x=10$, (a) $r=2$ and (b) $r=0.5$. We set
$k_BT/\hbar\omega_c=100$, $\alpha=0.1$. }\label{fig:1}
\end{figure}
In Fig. \ref{fig:1} we plot the time evolution of the EoF
calculated using the secular approximated solution and using the
exact solution in the regime $x \gg 1$ for two different initial
TWB states. For both initial conditions the secular approximation fails.
Remarkably, the exact solution containing the nonsecular terms
predicts a much longer disentanglement time. Furthermore the
difference in the disentanglement time predicted by the exact and
secular results increases for increasing values of initial
entanglement, i.e., for larger values of $r$. This result is
qualitatively independent of the analytic form of the reservoir
spectrum, as we will see in Sec. \ref{sec:secres} were the effect of
different spectra is considered.
\par
\begin{figure}[h!]
\begin{center}
 \includegraphics[width=0.49\columnwidth]{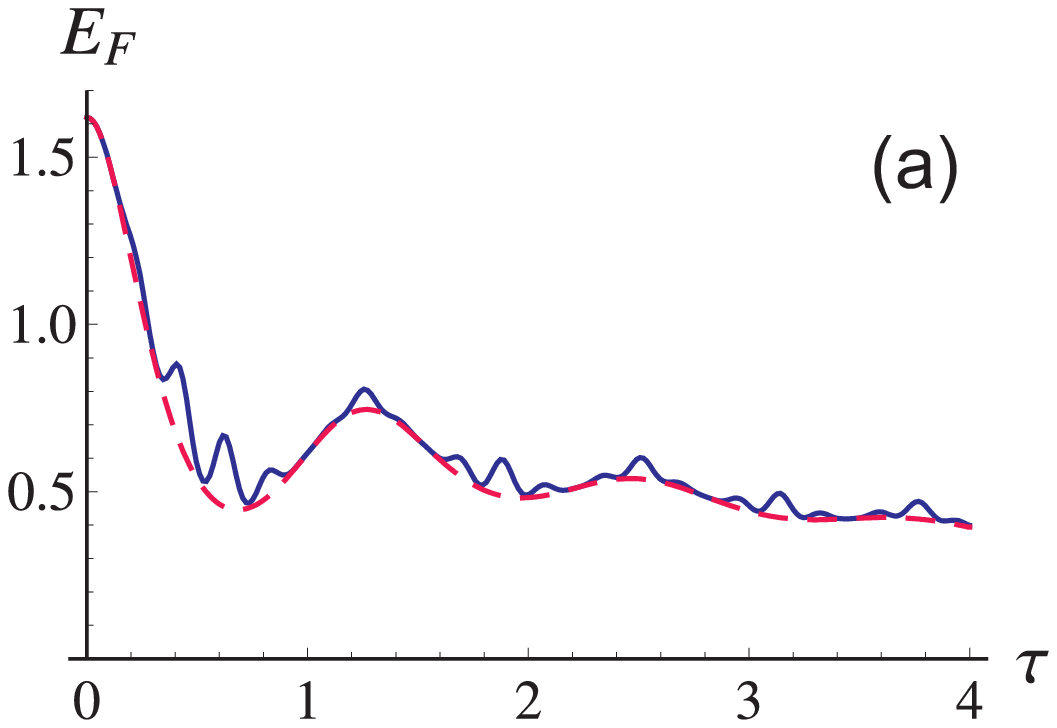}
 \includegraphics[width=0.49\columnwidth]{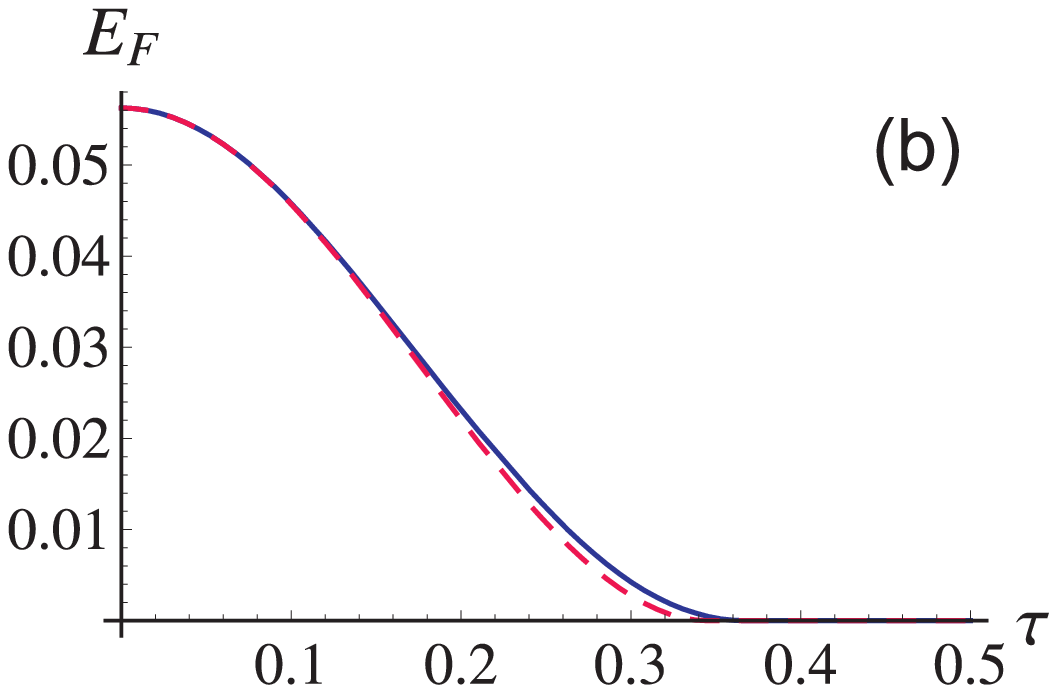}
\end{center}
\vspace{-0.1cm} \caption{(Colors online) Comparison between the
exact EoF dynamics (solid blue line) and the secular approximate
dynamics (dashed red line) as a function of $\tau=\omega_c t$, with
$x=0.2$, (a) $r=1$ and (b) $r=0.1$. We set
$k_BT/\hbar\omega_c=100$, $\alpha=0.1$.} \label{fig:2}
\end{figure}

For intermediate values of the parameter $x$, $x \le 1$, we observe
a stronger dependence on the initial value of entanglement. In Fig.
\ref{fig:2}, indeed, we see that for $x=0.2$ and $r=0.1$ (small
initial entanglement) the secular approximation works well, but for
higher values of the initial entanglement, $r=1$, nonsecular
oscillations, absent in the secular approximated solution, are
clearly visible in the exact dynamics.
\par
For $x \ll 1$, finally, the nonsecular oscillations decrease in
amplitude as the effective coupling with the environment decreases
and the secular coarse-grained solution describes well the dynamics
of the entanglement in the short non-Markovian time-scale,
independently from the initial condition.
This behavior is in agreement with the results of Ref.
\cite{QBMweak} where the weak coupling limit of the Master equation
for quantum Brownian motion is discussed. In particular, in Ref.
\cite{QBMweak} it is shown that, in the high-temperature and weak
coupling limits, the secular approximated Master equation is
accurate only in the regime $x \ll 1$, while for the other regimes
the system behaves as if it were subjected to a squeezed reservoir.
\par
Summarizing, for high-$T$ Ohmic reservoirs, the secular
approximation holds only in the regime $x \ll 1$. This result is
also valid for the sub-Ohmic and super-Ohmic environments.
\subsection{Zero-temperature regime}
From previous studies on open quantum systems interacting with
zero-temperature reservoirs we expect on the one hand a slower loss of
entanglement \cite{PraBec2004} and on the other hand more pronounced
non-Markovian features \cite{Alicki}, with respect to the $T\neq 0$
case. We will have a closer look at these general features of the
dynamics in Sec. \ref{sec:ED} and focus here on the validity of the
secular approximation.
\par
\begin{figure}[h!]
\begin{center}
\includegraphics[width=7.6cm]{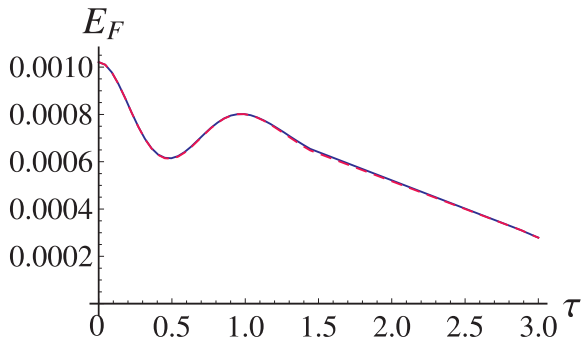}
\end{center}
\caption{(Colors online) The exact (blue solid line) and
the secular approximated dynamics (red dashed line) of the
$E_F$ vs $\tau=\omega_c t$ in a super-Ohmic reservoir at zero
temperature for $\alpha=0.1$, $r=0.01$ and $x=0.3$. The two
curves almost overlap perfectly.}\label{fig:3}
\end{figure}

We consider as an example  a super-Ohmic reservoir with $x=0.3$ and
look at the dynamics of a TWB with a small amount of initial
entanglement, $r=0.01$. As shown in Fig. \ref{fig:3}, the exact and
the secular approximated dynamics sensibly agree in this situation.
We have carefully examined the dynamical behavior for other values
of $x$ and of the initial squeezing parameter $r$ reaching the
conclusion that this is quite a general property of the system.
Therefore, in the description of bimodal CV quantum systems
interacting with zero-$T$ reservoirs, the secular approximation can
always be preformed and the effect of the nonsecular terms is always
negligible. This is a consequence of the fact that the secular terms
\eqref{SecCoeff} are temperature dependent through the diffusion
coefficients $\Delta(t)$ and $\Pi(t)$, and at $T=0$ their
contribution is rather small.
\subsection{Dependency on the reservoir spectrum} \label{sec:secres}
To conclude our analysis of the secular approximation we look at the
discrepancy between the secular and exact solutions for Ohmic,
sub-Ohmic and super Ohmic reservoirs. Since in the zero-$T$ case the
secular approximation always works well, we focus on the high-$T$
case and in particular on the $x \gg 1$ regime, where the
differences in the dynamics of the EoF are most pronounced.
\par
\begin{figure}[h!]
\begin{center}
 \includegraphics[width=0.49\columnwidth]{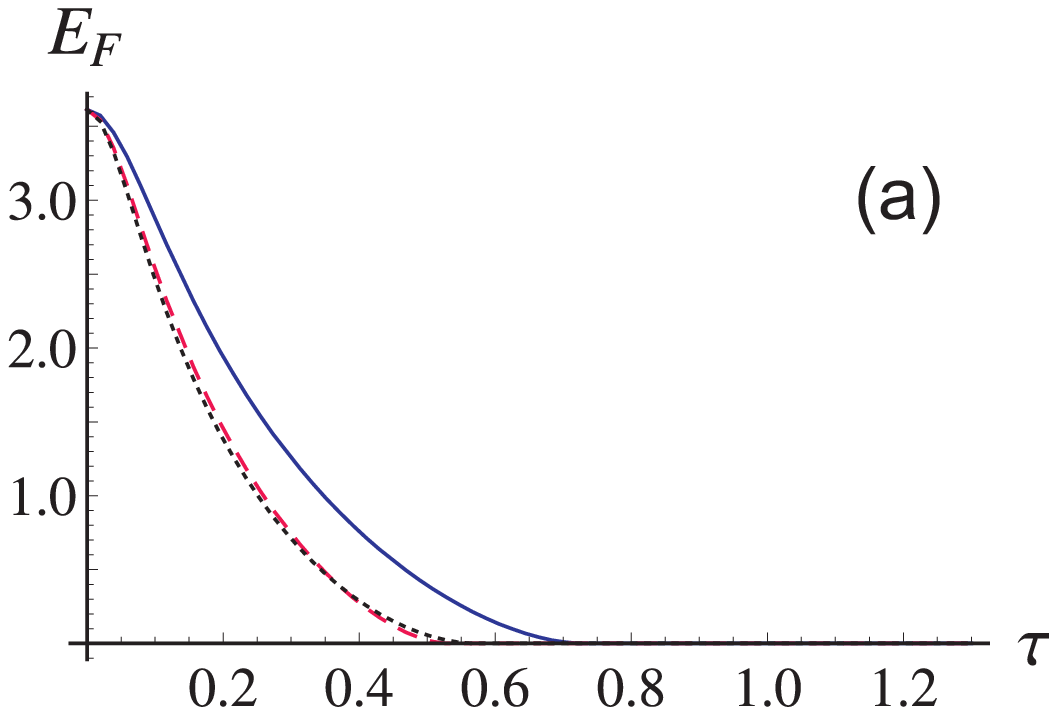}
 \includegraphics[width=0.49\columnwidth]{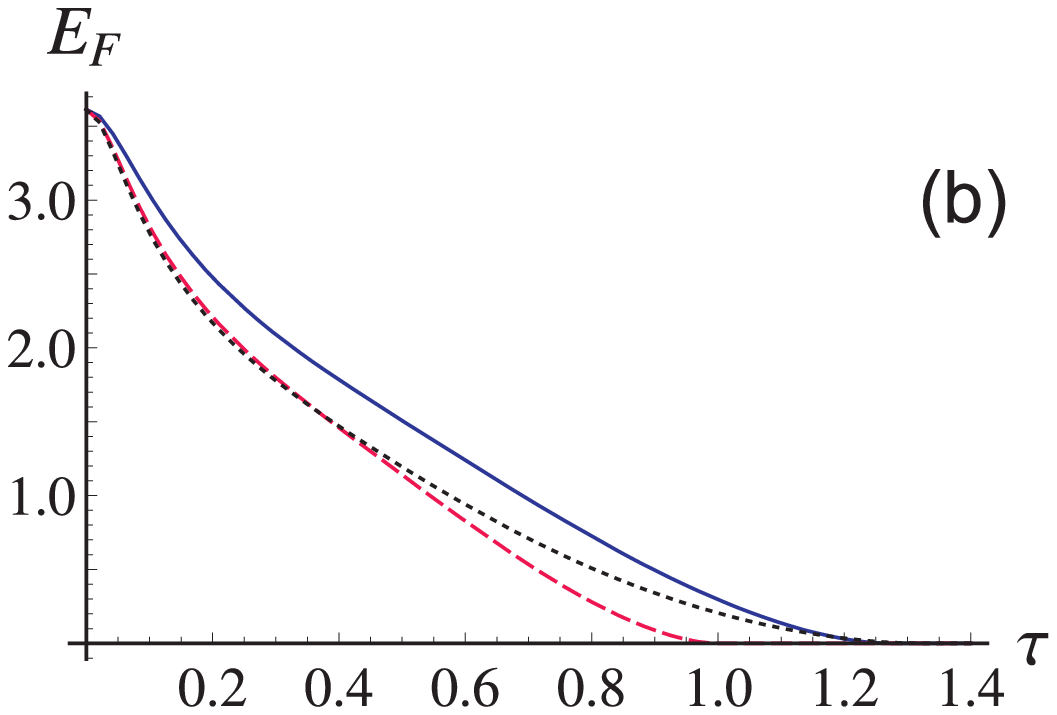}
\end{center}
\vspace{-0.1cm} \caption{(Colors online) Comparison between the
dynamics of $E_F$ for the Ohmic (blue solid line),
sub-Ohmic (red dashed line) and super-Ohmic (black dotted line)
reservoir spectra using (a) the secular approximated solution
and (b) the exact solution in the high-temperature limit with
$k_BT/\hbar\omega_c=100$, $\alpha=0.1$, $r=2$ and $x=10$.} \label{fig:4}
\end{figure}

In Fig. \ref{fig:4} we compare the dynamics of $E_F$ for the Ohmic,
sub-Ohmic and super-Ohmic reservoirs as given by the secular
approximation, Fig. \ref{fig:4} (a), with the exact case, Fig.
\ref{fig:4} (b), for $x=10$ and $r=2$. Comparing the two figures one
clearly sees that the secular approximation does not affect the
dynamics in an equal way for the three different spectral
distributions. The exact disentanglement time is almost tripled for
the super-Ohmic environment and doubled for the Ohmic case. The
sub-Ohmic case is less affected. In all cases, however, and for all
values of initial entanglement, the exact calculation predicts a
longer survival time of entanglement.
\section{Entanglement dynamics} \label{sec:ED}
\subsection{General features: Three different dynamical regimes}
In discrete variable quantum systems the phenomenon of ESD
has recently received a lot of attention \cite{YuScience}.
In that context the basic system studied consists
of two qubits interacting with either independent or common
reservoirs.  An exact solution has been derived both for independent
\cite{Bellomo}  and for common reservoirs \cite{Mazzola},
and it has been shown that revivals of entanglement
due to the reservoir memory may occur after an initial sudden death
interval. The common reservoir scenario is generally characterized
by a non-zero long time entanglement due to both the
reservoir-mediated interaction between the qubits and the existence
of a decoherence free subradiant state acting as an
entanglement-trap \cite{Francesco1,Francesco2,Kari}. In the
independent reservoir case, on the contrary, in the long time limit
one always observes complete disentanglement.
\par
In the following we focus on the case in which the carriers of quantum
information are continuous, instead of discrete, variable quantum
systems. While the theoretical treatments of CV quantum channels is
fundamentally different from the one of discrete channels, we find
that some similarities in the entanglement dynamics do exist. In
particular, in the common reservoir scenario, due to the
environment-mediated interaction between the two CV channels, the
asymptotic long time entanglement maybe non-zero, even for high-$T$
reservoirs \cite{Paz08,Paz09}. Moreover, non-Markovian studies show
the occurrence of revivals of entanglement both in the common and in
the independent reservoir cases \cite{Paz08,Paz09,MaOliPa}.
\par
\begin{figure}[h!]
\begin{center}
 \includegraphics[width=0.49\columnwidth]{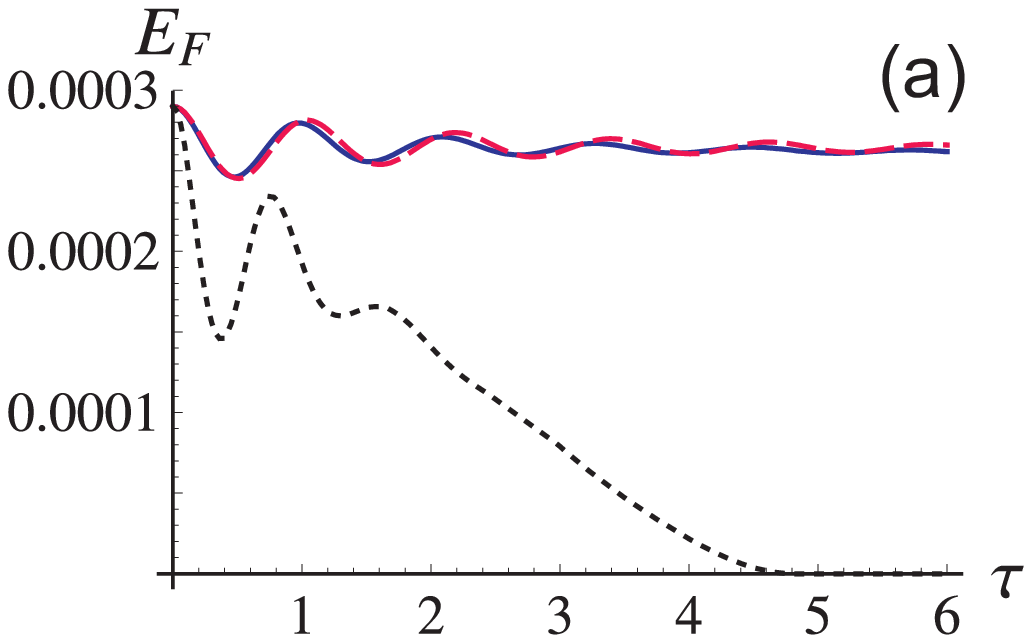}
 \includegraphics[width=0.49\columnwidth]{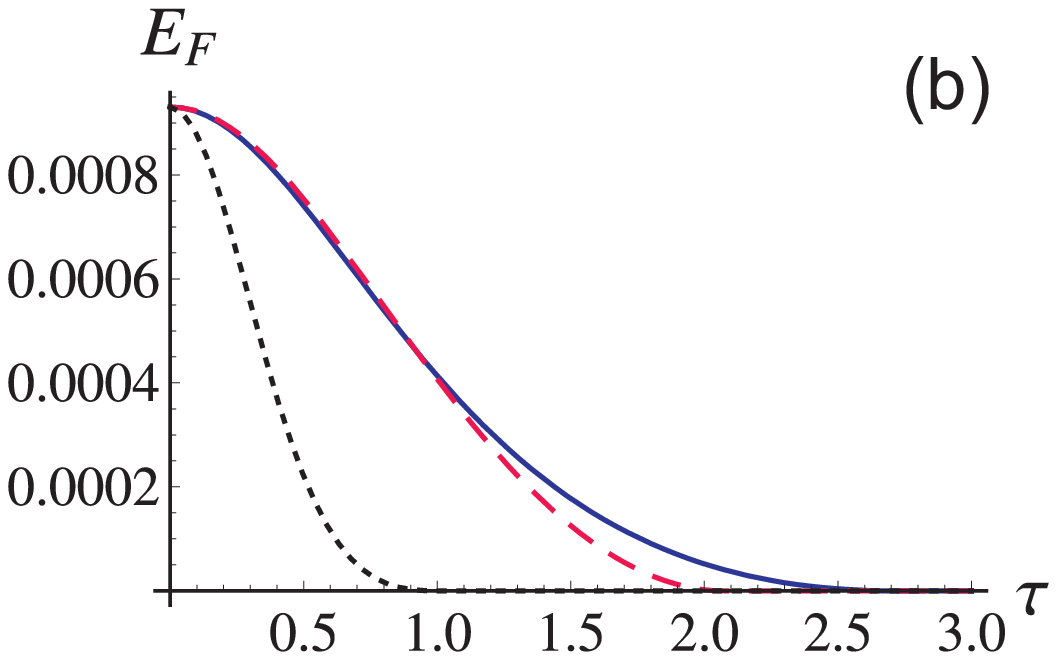}
\end{center}
\vspace{-0.1cm} \caption{(Colors online) Dynamics of $E_F$ at zero
temperature for $\alpha=0.1$ in the case of Ohmic (blue line),
sub-Ohmic (red line) and super-Ohmic (black line) reservoirs for
$x=0.2$ and $r=0.005$ (a) and for $x=10$ and $r=0.01$ (b).} \label{fig:5}
\end{figure}
\begin{figure}[h!]
\begin{center}
 \includegraphics[width=0.9\columnwidth]{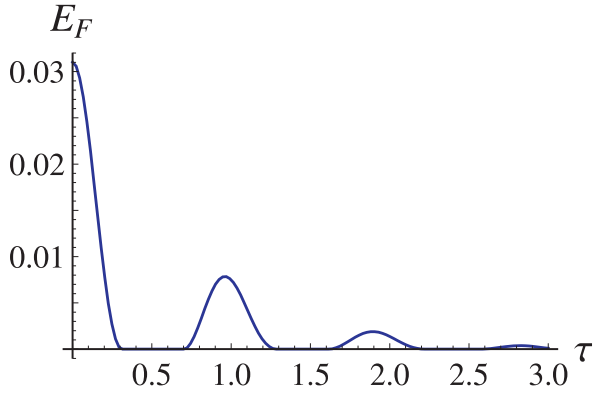}
\end{center}
\vspace{-0.1cm} \caption{(Colors online) $E_F$ dynamics vs
$\tau=\omega_c t$ for an high-$T$ Ohmic reservoir with
$k_BT/\hbar\omega_c=100$,$\alpha=0.1$, $x=0.15$ and $r=0.06$. Note
the presence of entanglement revivals due to negative values of the
Master equation coefficients.} \label{fig:6}
\end{figure}

Here we show that, for independent reservoirs and for $x \gg 1$, the
phenomenon of ESD occurs both in the high-$T$ and, for $r\ll 1$, in
the zero-$T$ cases,  independently from the reservoirs spectra, as
one can see from Fig. \ref{fig:1}, Fig. \ref{fig:4} and Fig.
\ref{fig:5} (b).
The  ESD regime can directly be linked to the behavior of the
time-dependent coefficients appearing in the Master equation
\eqref{HuPaZang}. For high-$T$ and $x \gg 1$, indeed, independently
from the reservoir spectra, the time dependent coefficients are
always positive at every time instant \cite{Paa}. In systems
described by time-convolutionless Master equations non-Markovian
features typically occur when the time-dependent coefficients
temporarily attain negative values \cite{Piilo,PiiloPRA}. When this
happens revivals of entanglement may occur since the system restores
partially the quantum coherence previously lost due to the
interaction with the environment. An example of non-Markovian
revivals due to the reservoir memory effects, and therefore
connected to the negativity of the time dependent coefficients, is
shown  in Fig. \ref{fig:6}, where we plot the EoF dynamics for an
Ohmic reservoir in the high-T limit. In general non-Markovian
revivals of entanglement occur for $x \ll 1$. In this case we have
seen that the secular approximation works well and we know from
previous studies (See Ref. \cite{Paa,PRAnonL}) that the time
dependent coefficients attain negative values for all reservoir
spectra.
\par
\begin{figure}[h!]
\begin{center}
\includegraphics[width=0.99\columnwidth]{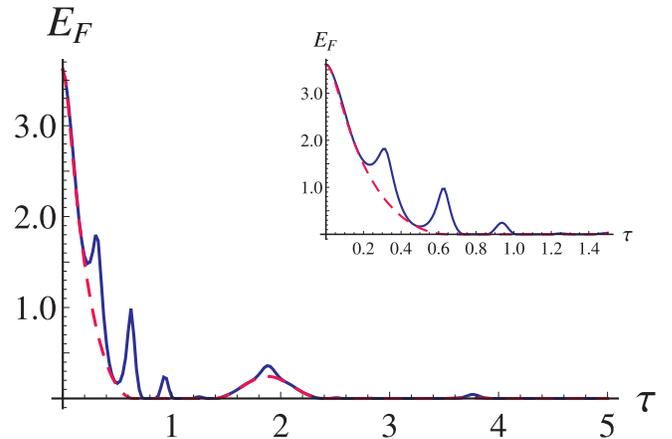}
\end{center}
\caption{(Colors online) $E_F$ for a Sub-Ohmic reservoir in the
high-T limit ($k_BT/\hbar\omega_c=100$) with $\alpha=0.1$, $r=2$ and
$x=0.3$. The solid blue line is the exact solution while the dashed
red line is the secular solution. The inset is a magnification of
the temporal region $0\leq\tau\leq 1.5$.}\label{fig:7}
\end{figure}

Our exact approach  allows us, moreover, to single out another type
of revivals, not related directly to the negativity of the
time-dependent coefficients, and therefore on the reservoir memory,
but rather on the presence of nonsecular terms. An example of the
nonsecular revivals is given in Fig. \ref{fig:7}, where the exact
and the secular solutions are compared. In this case the
time-dependent coefficients are positive for $\tau \lesssim 1.4$ and
the partial and temporary restoration of entanglement in that time
interval is due to the presence of the counter-rotating terms in the
microscopic Hamiltonian model. It is often believed that the
nonsecular or counter-rotating terms significantly affect the
dynamics only in the strong coupling limit. For discrete variable
systems indeed, a very recent study has shown the non-negligible
effect of nonsecular terms in the strong coupling limit
\cite{Fizek}. In this case the authors show that the exact dynamics
causes a faster loss of entanglement with respect to the secular
case. Here we show that also in the weak coupling limit these terms
give a non-negligible contribution in the short non-Markovian time
scale.
In general, for intermediate values of $x$, the dynamics will
display both non-secular and non-Markovian revivals. Indeed, if we
look at the dynamics for $\tau \gtrsim 1.4$ we see that a
non-Markovian revival, with a superimposed nonsecular revival,
occurs at $1.5 \lesssim \tau \lesssim 2.5$, as shown in Fig.
\ref{fig:7} (b). The border between the NMRev and the NSRev
dynamical regimes is therefore a blurred region in which both
effects occur at the same time. In this case both the reservoir
memory and the nonsecular terms contribute to the reappearance of
previously lost entanglement.
\par
A more detailed discussion is required to explain the presence of ESD in
the zero-temperature cases shown in \ref{fig:5} (a)-(b). The Markovian
theory of two-mode continuous variable channels predicts, for both
the common and the independent reservoirs, the existence of a finite
time of disentanglement for an initial TWB state when $T>0$
\cite{PraBec2004}. In the independent reservoirs model at $T=0$,
however, the Markovian disentanglement time is infinite (no ESD).
Since the exact Master equation \eqref{HuPaZang}  coincides with the
approximate Born-Markov Master equation for weak couplings and for times longer
than the reservoirs correlation time, one would expect our
non-Markovian model to give the same prediction for the
disentanglement time than the Markovian one. Stated another way, one
would not expect ESD.
\par
However, one should keep in mind that the Markovian approximation is
always a coarse graining in time and therefore it does not allow us
to predict the short time non-Markovian behavior. If at short times
the initial entanglement is lost and no non-Markovian revivals
occur, entanglement cannot reappear at longer times. Consequently
the entanglement will remain zero also in the asymptotic Markovian
long-time region. This is exactly what may happen when the initial
amount of entanglement is small ($r \ll 1$). In this case, indeed,
for some reservoirs spectra and values of $x$, the short time
non-Markovian dynamics shows the occurrence of sudden death of
entanglement. Since the state remains separable for times greater
than the reservoirs correlation time, entanglement revivals cannot
appear.
\par
For higher values of initial entanglement, on the other hand, the
exact non-Markovian theory does not lead to a sudden death in the
short time scale, thus the state is still entangled when reaching the
Markovian time-region and therefore the Markovian prediction of an
infinite disentanglement time at $T=0$ still holds.
\par
Summarizing, for $x \gg 1$  ESD occurs, independently from the
reservoir spectrum,  both in high $T$ reservoirs (for all values of
$r$) and in zero-$T$ reservoirs (for $r \ll 1$). When $x \ll 1$ we
are generally in the NMRev region, independently from the reservoir
spectrum. One should note, however, that if the initial entanglement
is very small ($r\ll 0.1$) entanglement oscillations do not have time to
take place and only ESD is observed. For intermediate values of $x$
the dynamical regimes strongly depend both on the reservoir spectrum
and on the initial entanglement. More specifically, for high-$T$
reservoirs one can have any of the three ESD, NMRev and NSRev
behaviors, as well as a combination of NMRev and NSRev. For zero-$T$
reservoirs the ESD or NMRev regimes exist only if the initial amount
of entanglement is small ($r\ll1$), for other initial values of $r$
entanglement is never lost, in accordance with the Markovian theory.
\subsection{Comparative study of Ohmic, sub-Ohmic and super-Ohmic
reservoirs}
In this section we investigate the differences in the loss of
entanglement due to different reservoir spectra. Different physical
systems are characterized by different environmental spectral
densities, e.g., it is well known that solid-state systems are
subjected to sub-Ohmic  $1/f$ noise. Such a comparative study,
hence, allows to understand which physical context is more \lq
quantum information friendly\rq, in the sense of allowing
entanglement to live longer.
\par
\begin{figure}[h!]
\begin{center}
  \includegraphics[width=0.99\columnwidth]{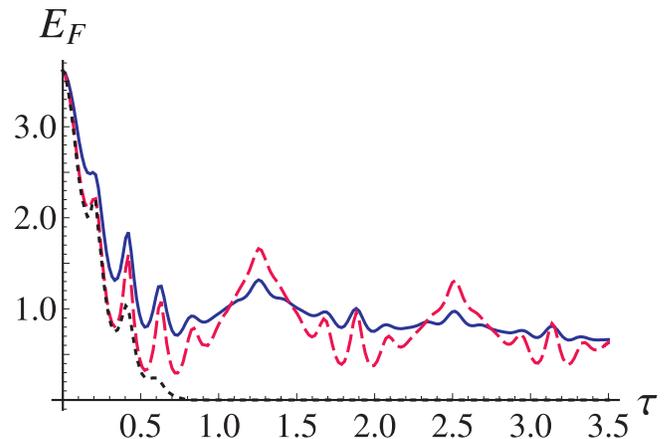}
\end{center}
\vspace{-0.1cm} \caption{(Colors online) Dynamics of the $E_F$ in
the high-T temperature limit $k_BT/\hbar\omega_c=100$ for $r=2$ and
$x=0.2$ in the case of Ohmic (blue solid line), sub-Ohmic (red
dashed line) and super-Ohmic (black dotted line) environments.}
\label{fig:8}
\end{figure}

We begin considering the high-$T$ reservoir case. In Fig.
\ref{fig:4} (b) we have seen that in the ESD regime and for high
temperatures the behavior of entanglement is qualitatively similar.
The disentanglement time is not strongly dependent on the reservoir
spectrum and the sub-Ohmic environment displays the faster loss of
entanglement. For intermediate values of $x$, however, the time
evolution of $E_F$ shows a much richer behavior and a much stronger
dependence on the form of the spectrum, as one can see in Fig.
\ref{fig:8}. In this case the super-Ohmic environment shows a much
faster loss of entanglement than the Ohmic and sub-Ohmic. In the
Ohmic and sub-Ohmic cases one can clearly see the nonsecular
oscillations superimposed to the non-Markovian oscillations in the
dynamics of $E_F$, the latter ones having longer period and larger
amplitude.
\par
The entanglement dynamics for zero-$T$ reservoirs is plotted in Fig.
\ref{fig:5} for (a) $x=0.2$ and (b) $x=10$. Also in this case the
super-Ohmic reservoir shows a much faster entanglement loss than the
sub-Ohmic and the Ohmic ones. This is especially evident in the
intermediate $x$ region of Fig.  \ref{fig:5} (a). The Markovian and
RW theory of entanglement dynamics for TWBs in  independent
reservoirs predicts that the disentanglement time (or separability
time) should approach infinity for $T \rightarrow 0$
\cite{PraBec2004}. Our results demonstrate that this conclusion, in
the case of small initial entanglement, is a consequence of the
Markovian approximation and that the exact non-Markovian theory
predicts that, even for weak system-reservoir coupling, in the $x
\gg 1$ region the entanglement survives only for a short time. For
intermediate values of $x$, however, the disentanglement time
approaches the Markovian prediction for the sub-Ohmic and Ohmic
cases. Indeed we see that after initial non-Markovian oscillations
$E_F$ approaches its stationary non-zero Markovian value.
\section{Summary and Conclusions}\label{s:concl}
In this paper we have studied a bimodal CV quantum system interacting
with independent structured reservoirs in thermal equilibrium. We focus
on the dynamics of the entanglement, as measured by the entanglement of
formation, for the two modes initially excited in a twin-beam state and
for different reservoir spectral distributions (Ohmic, sub-Ohmic and
super-Ohmic). Under the only assumption of weak coupling we have
obtained an exact analytic solution for the time-dependent two-mode
covariance matrix describing the state of our system in the short time
non-Markovian limit.
\par
In the first part of the paper we unravel the role of the secular
approximation in our specific system and in particular in the context of
the entanglement dynamics. By comparing the exact solution with the
solution in the secular assumption we found that, in general, the high-T
dynamics is affected by this approximation, while the $T=0$ case is not.
More specifically at high temperatures we have shown that for $x\gg1$
both the exact and secular solutions predict the occurrence of ESD.
However the entanglement persists longer in the exact solution. On the
contrary in the limit of $x\ll 1$ the solutions agree. These two results
are independent from the analytic expression of the reservoirs spectral
distributions and from the initial amount of the entanglement. For
intermediate values of $x$ the situation is more complicated, and the
validity of the secular approximation strongly depends on the expression
of the spectral distribution and on the initial amount of entanglement.
\par
The second aim of the paper was the investigation of the
entanglement dynamics as a function of the reservoir spectrum, the
temperature and the initial amount of entanglement. Essentially we
observed the presence of three different behaviors: sudden death of
entanglement (ESD), non-Markovian entanglement revivals and oscillations
(NMRev) and oscillations or revivals related only to the
secular coefficients (NSRev).
\par
At high temperatures ESD appears for each
value of the initial entanglement. Moreover, for $x\gg 1$ there are no
revivals while they are typical of the dynamics for $x\ll 1$. Because in this
limit the secular and exact dynamics almost coincide, these
revivals are due to the negativity of the Master equation
coefficients (NMRev). For intermediate values of $x$,  the time evolution
strongly depends on the initial amount of entanglement. If the entanglement is
small ($r<1$) only ESD is observed. For larger value of $r$, in
general, the behavior is characterized by oscillations and revivals.
Some of these revivals occur in correspondence of positive value
of the time-dependent coefficients. Therefore they exist as a
consequence of the secular coefficients only (NSRev).
\par
The situation for $T=0$ is characterized by a slower rate of
entanglement deterioration. Therefore, for short times, entanglement
sudden death and revivals can be observed only for very small initial
entanglement ($r\ll 1$). In these cases, and when $x\gg 1$, ESD exists
independently from the reservoir spectrum while for $x\ll 1$ EDS is
present in the super-Ohmic case only.  Hence, the asymptotic long time
Markovian dynamics of entanglement, and therefore also the Markovian
prediction about the disentanglement time, may be strongly affected by
the non-Markovian short time correlations. When this happens, the
non-Markovian theory predicts a finite disentanglement time in contrast
to the Markovian prediction. When $r>0.1$  the short time non-Markovian
dynamics is characterized by oscillations only, the EoF remaining
positive. Therefore, for long times the Markovian prediction of an
infinite disentanglement time is recovered.
\par
In recent years there have been a lot of interest in the entanglement
dynamics in CV quantum channels, both for common and independent
reservoirs. Our work finds its  place in this context as an attempt to
investigate the non-Markovian short time dynamics of entanglement in
different physical scenarios. We believe that our results, showing the
effects of different reservoirs on the time evolution of entanglement in
CV quantum channels, will pave the way to the implementation of
engineered reservoir control schemes as the one recently reported in
Ref. \cite{Itano09} for qubits.
\par
\acknowledgements
SM and RV acknowledge financial support from the Academy of Finland
(Projects No.~115682,and No.~8125004), the V\"ais\"al\"a Foundation, the
Magnus Ehrnrooth Foundation, the Emil Aaltonen Foundation and the Turku
Collegium of Science and Medicine. This work has been partially
supported by the CNR-CNISM convention.
\begin{widetext}
\appendix*
\section{Time dependent coefficients at the second order in $\alpha$}
Here we provide the exact analytic expressions of the time-dependent
coefficients of the Master Equation given in \eqref{Coeff1}. In each
subsection we consider a single reservoir spectral function and
evaluate the temperature independent damping coefficient
$\gamma(t)$, the diffusion coefficients in the high temperature
regime $\Delta_T(t)$ and $\Pi_T(t)$, and the diffusion coefficients
at $T=0$, $\Delta_0(t)$ and $\Pi_0(t)$. The expression for
$\Gamma(t)$, $\Delta_{\Gamma}(t)$ and the secular terms
\eqref{SecCoeff} follow through. We made use of the following
special mathematical functions \cite{Abramovitz}
\begin{equation}\begin{split}
&Ei(z)=-\int_{-z}^{+\infty}\frac{e^{-t}}{t}dt\qquad\qquad
Ci(z)=-\int_z^{+\infty}\frac{\cos t}{t}dt\qquad\qquad
Si(z)=\int_0^z\frac{\sin t}{t}dt\\
&Sih(z)=\int_0^z\frac{\sinh t}{t}dt\qquad\qquad
Erf(z)=\frac{2}{\sqrt{\pi}}\int_0^ze^{-t^2}dt\nonumber
\end{split}\end{equation}
\subsection{Ohmic Reservoir  $s=1$}
\begin{equation}\begin{split}
\gamma(t)=\frac{\omegaz\alpha^2}{4}\biggl\{ie^{-1/x}
\biggl[Ei\biggl(\frac{1-i\tau}{x}\biggl)-Ei\biggl(\frac{1+i\tau}{x}\biggl)\biggl]+e^{1/x}\biggl[
2\pi+iEi\biggl(\frac{i\tau-1}{x}\biggl)-iEi\biggl(-\frac{1+i\tau}{x}\biggl)\biggl]-
\frac{4x\sin(\tau/x)}{1+\tau^2}\biggl\}
\end{split}\end{equation}

\begin{equation}\begin{split}
\Delta_T(t)=-\frac{k_BT\alpha^2}{\hbar}\biggl\{
i\cosh\biggl(\frac{1}{x}\biggl)\biggl[Ci\biggl(\frac{\tau-i}{x}\biggl)-Ci\biggl(\frac{\tau+i}{x}\biggl)-i\pi
\biggl]+\sinh\biggl(\frac{1}{x}\biggl)\biggl[Si\biggl(\frac{\tau-i}{x}\biggl)+
Si\biggl(\frac{\tau+i}{x}\biggl)\biggl]\biggl\}
\end{split}\end{equation}

\begin{equation}\begin{split}
&\Pi_T(t)=\frac{k_BT\alpha^2}{\hbar}
\biggl\{\sinh\biggl(\frac{1}{x}\biggl)
\biggl[Ci\biggl(\frac{\tau-i}{x}\biggl)+Ci\biggl(\frac{\tau+i}{x}\biggl)
-Ci\biggl(-\frac{i}{x}\biggl)-Ci\biggl(\frac{i}{x}\biggl)\biggl]+
\cosh\biggl(\frac{1}{x}\biggl)\\
&\times\biggl[2Sih\biggl(\frac{1}{x}\biggl)
-iSi\biggl(\frac{\tau-i}{x}\biggl)+iSi\biggl(\frac{\tau+i}{x}\biggl)\biggl]\biggl\}
\end{split}\end{equation}

\begin{equation}\begin{split}
\Delta_0(t)=\frac{\omegaz\alpha^2}{4}\biggl\{ie^{-1/x}
\biggl[Ei\biggl(\frac{1-i\tau}{x}\biggl)-Ei\biggl(\frac{1+i\tau}{x}\biggl)\biggl]-e^{1/x}\biggl[
2\pi+iEi\biggl(\frac{i\tau-1}{x}\biggl)-iEi\biggl(-\frac{1+i\tau}{x}\biggl)\biggl]+
\frac{4x\tau\cos(\tau/x)}{1+\tau^2}\biggl\}
\end{split}\end{equation}

\begin{equation}\begin{split}
&\Pi_0(t)=\frac{\omegaz\alpha^2}{4}\biggl\{-e^{-1/x}
\biggl[Ei\biggl(\frac{1-i\tau}{x}\biggl)+Ei\biggl(\frac{1+i\tau}{x}\biggl)
-2Ei\biggl(\frac{1}{x}\biggl)\biggl]+e^{1/x}\\
&\times\biggl[
2Ei\biggl(-\frac{1}{x}\biggl)-Ei\biggl(\frac{i\tau-1}{x}\biggl)-
Ei\biggl(-\frac{1+i\tau}{x}\biggl)\biggl]+
\frac{4x\tau\sin(\tau/x)}{1+\tau^2}\biggl\}
\end{split}\end{equation}

\subsection{Sub-Ohmic Reservoir $s=1/2$}
\begin{equation}\begin{split}
\gamma(t)&=\frac{\alpha^2\omega_0\sqrt{\pi}}{4}
\biggl\{\frac{2ix\sin(t/x)(1+it+\sqrt{1+t^2})}{\sqrt{1-it}(t-i)}+e^{-1/x}\sqrt{\pi
x}\biggl[Erf\biggl((-1)^{3/4}\sqrt{\frac{-i+t}{x}}\biggl)\\
&-ie^{2/x}Erf\biggl((-1)^{3/4}\sqrt{\frac{-i+t}{x}}\biggl)+
ie^{2/x}Erf\biggl((-1)^{3/4}\sqrt{\frac{i+t}{x}}\biggl)+Erf
\biggl((-1)^{1/4}\sqrt{\frac{i+t}{x}}\biggl)\biggl]\biggl\}
\end{split}\end{equation}

\begin{equation}\begin{split}
\Delta_T(t)&=-\frac{\alpha^2\pi
k_BT}{2\hbar}\sqrt{x}e^{-1/x}\biggl\{Erf\biggl((-1)^{1/4}\sqrt{\frac{i-t}{x}}\biggl)
-Erf\biggl((-1)^{1/4}\sqrt{\frac{i+t}{x}}\biggl)\\
&+ie^{2/x} \biggl[Erf\biggl((-1)^{3/4}\sqrt{\frac{i+t}{x}}\biggl)
-Erf\biggl((-1)^{3/4}\sqrt{\frac{i-t}{x}}\biggl)\biggl]\biggl\}
\end{split}\end{equation}

\begin{equation}\begin{split}
\Pi_T(t)&=\frac{\alpha^2\pi
k_BT}{2\hbar}\sqrt{x}e^{-1/x}\biggl\{e^{2/x}\biggl[Erf\biggl((-1)^{3/4}\sqrt{\frac{i-t}{x}}\biggl)
-Erf\biggl((-1)^{3/4}\sqrt{\frac{i+t}{x}}\biggl)-2Erf\biggl(\sqrt{\frac{1}{x}}\biggl)\biggl]\\
&+i\biggl[Erf\biggl((-1)^{1/4}\sqrt{\frac{i-t}{x}}\biggl)+Erf\biggl((-1)^{1/4}\sqrt{\frac{i+t}{x}}\biggl)
-2Erf\biggl(i\sqrt{\frac{1}{x}}\biggl)\biggl]\biggl\}
\end{split}\end{equation}

\begin{equation}\begin{split}
\Delta_0(t)&=\frac{\alpha^2\omega_0\sqrt{\pi}}{4}\biggl\{\frac{2ix\cos(t/x)}{\sqrt{1+t^2}}(\sqrt{1-it}+\sqrt{1+it})+
e^{-1/x}\sqrt{x\pi}\biggl[-Erf\biggl((-1)^{3/4}\sqrt{\frac{-i+t}{x}}\biggl)\\
&+Erf\biggl((-1)^{1/4}\sqrt{\frac{i+t}{x}}\biggl)+ie^{2/x}Erf\biggl((-1)^{1/4}\sqrt{\frac{-i+t}{x}}\biggl)+
ie^{2/x}Erf\biggl((-1)^{3/4}\sqrt{\frac{i+t}{x}}\biggl)\biggl]
\biggl\}
\end{split}\end{equation}

\begin{equation}\begin{split}
\Pi_0(t)&=\frac{\alpha^2\omega_0\sqrt{\pi}}{4}\biggl\{-\frac{2ix\sin(t/x)}{\sqrt{1+t^2}}(\sqrt{1-it}+\sqrt{1+it})+e^{1/x}\sqrt{\pi
x}\biggl[2e^{2/x}Erf\biggl(\sqrt{\frac{1}{x}}\biggl)-e^{2/x}Erf\biggl((-1)^{1/4}\sqrt{\frac{-i+t}{x}}\biggl)\\
&+e^{2/x}Erf\biggl((-1)^{3/4}\sqrt{\frac{i+t}{x}}\biggl)+iErf\biggl((-1)^{3/4}\sqrt{\frac{-i+t}{x}}\biggl)
+iErf\biggl((-1)^{1/4}\sqrt{\frac{i+t}{x}}\biggl)-2iErf\biggl(i\sqrt{\frac{1}{x}}\biggl)\biggl]
\biggl\}
\end{split}\end{equation}

\subsection{Super-Ohmic Reservoir  $s=3$}

\begin{equation}\begin{split}
\gamma(t)&=\frac{\alpha^2\omega_0}{4
x^2(1+t^2)^3}\biggl\{8x^2(1+t^2)t\cos\biggl(\frac{t}{x}\biggl)+4x[-(1+t^2)^2+2(3t^2-1)x^2]\sin\biggl(\frac{t}{x}\biggl)+e^{-1/x}(1+t^2)^3
\biggl[2e^{2/x}\pi\\
&+iEi\biggl(\frac{1-it}{x}\biggl)
-ie^{2/x}Ei\biggl(-\frac{1+it}{x}\biggl)-iEi\biggl(\frac{1+it}{x}\biggl)+ie^{2/x}Ei\biggl(-\frac{1-it}{x}\biggl)\biggl]\biggl\}
\end{split}\end{equation}

\begin{equation}\begin{split}
\Delta_T(t)&=\frac{\alpha^2 k_BT}{2\hbar
x^2(1+t^2)^2}\biggl\{8x^2t\cos\biggl(\frac{t}{x}\biggl)-4(1+t^2)x\sin\biggl(\frac{t}{x}\biggl)+e^{-1/x}(1+t^2)^2
\biggl[2e^{2/x}\pi+iEi\biggl(\frac{1-it}{x}\biggl)\\
&-ie^{2/x}Ei\biggl(-\frac{1+it}{x}\biggl)-iEi\biggl(\frac{1+it}{x}\biggl)+ie^{2/x}Ei\biggl(-\frac{1-it}{x}\biggl)\biggl]\biggl\}
\end{split}\end{equation}

\begin{equation}\begin{split}
\Pi_T(t)&=\frac{\alpha^2 k_BT}{2\hbar
x^2(1+t^2)^2}\biggl\{4x(1+t^2)\cos\biggl(\frac{t}{x}\biggl)+8tx^2\sin\biggl(\frac{t}{x}\biggl)-e^{-1/x}(1+t^2)^2
\biggl[4e^{1/x}x+2e^{2/x}Ei\biggl(-\frac{1}{x}\biggl)-2Ei\biggl(\frac{1}{x}\biggl)\\
&+Ei\biggl(\frac{1-it}{x}\biggl)-e^{2/x}Ei\biggl(-\frac{1+it}{x}\biggl)+Ei\biggl(\frac{1+it}{x}\biggl)-e^{2/x}Ei\biggl(-\frac{1-it}{x}\biggl)\biggl]\biggl\}
\end{split}\end{equation}

\begin{equation}\begin{split}
\Delta_0(t)&=\frac{\alpha^2\omega_0}{2x^2}\biggl\{\frac{2x}{(1+t^2)^3}\biggl[-(1-t^4)x\sin\biggl(\frac{t}{x}\biggl)+
t\cos\biggl(\frac{t}{x}\biggl)\biggl(1+2t^2+t^4+6x^2-2x^2t^2\biggl)\biggl]\\
&+i\sinh\biggl(\frac{1}{x}\biggl)\biggl[Ci\biggl(\frac{-i+t}{x}\biggl)-Ci\biggl(\frac{i+t}{x}\biggl)+i\pi\biggl]+\cosh\biggl(\frac{1}{x}\biggl)\biggl[
Si\biggl(\frac{-i+t}{x}\biggl)+Si\biggl(\frac{i+t}{x}\biggl)\biggl]\biggl\}
\end{split}\end{equation}

\begin{equation}\begin{split}
\Pi_0(t)&=\frac{\alpha^2\omega_0}{2x^2}\biggl\{-2x^2+\frac{2x}{(1+t^2)^3}\biggl[(1-t^4)x\cos\biggl(\frac{t}{x}\biggl)+
t\sin\biggl(\frac{t}{x}\biggl)\biggl(1+2t^2+t^4+6x^2-2x^2t^2\biggl)\biggl]\\
&-\cosh\biggl(\frac{1}{x}\biggl)\biggl[Ci\biggl(\frac{-i+t}{x}\biggl)+Ci\biggl(\frac{i+t}{x}\biggl)-i\pi\biggl]+\sinh\biggl(\frac{1}{x}\biggl)\biggl[
-2Sih\biggl(\frac{1}{x}\biggl)+iSi\biggl(\frac{-i+t}{x}\biggl)-iSi\biggl(\frac{i+t}{x}\biggl)\biggl]\biggl\}
\end{split}\end{equation}

\end{widetext}


\end{document}